\documentclass[twocolumn,twoside]{IEEEtran}
\usepackage{mathpazo}
\usepackage{times}

\usepackage{amsmath}
\usepackage{amsfonts}
\usepackage{latexsym}
\usepackage{amssymb}

\usepackage{upref}
\usepackage{theorem}
\usepackage{graphicx}
\usepackage{psfrag}
\usepackage{cite}
\usepackage{color}

\renewcommand{\markboth}[2]
{\renewcommand{\leftmark}{#1}\renewcommand{\rightmark}{#2}}

\theoremstyle{plain}
\theorembodyfont{\normalfont\itshape}

\theoremstyle{plain}
\theorembodyfont{\normalfont\slshape}

\newtheorem{thm}{Theorem$\!$}
\newenvironment{theorem}
{\begin{thm}\hspace*{-1ex}{\bf.}}{\end{thm}}

\newtheorem{clm}[thm]{Claim$\!$}
\newenvironment{claim}{\begin{clm}\hspace*{-1ex}{\bf.}}{\end{clm}}

\newtheorem{lem}[thm]{Lemma$\!$}
\newenvironment{lemma}{\begin{lem}\hspace*{-1ex}{\bf.}}{\end{lem}}

\newtheorem{prop}[thm]{Proposition$\!$}

\newtheorem{cor}[thm]{Corollary$\!$}
\newenvironment{corollary}{\begin{cor}\hspace*{-1ex}{\bf.}}{\end{cor}}

\newtheorem{defn}[thm]{Definition$\!$}
\newenvironment{definition}{\begin{defn}\hspace*{-1ex}{\bf.}}{\end{defn}}

\newtheorem{xmpl}[thm]{Example$\!$}
\newenvironment{example}{\begin{xmpl}\hspace*{-1ex}{\bf.}}{\hfill$\Box$\end{xmpl}}

\newtheorem{cnstr}{Construction$\!$}
\newenvironment{construction}{\begin{cnstr}\hspace*{-1ex}{\bf.}}{\end{cnstr}}

\newcounter{enumrom}
\renewcommand{\theenumrom}{(\roman{enumrom})}

\makeatletter
\renewcommand{\@endtheorem}{\endtrivlist}
\makeatother

\makeatletter
\renewcommand{\thefigure}{{\@arabic\c@figure}}
\renewcommand{\fnum@figure}{{\bf Figure\,\thefigure}}
\makeatother

\newcommand{\cC}{\mathcal{C}}
\newcommand{\cD}{\mathcal{D}}

\newcommand{\cQ}{\mathcal{Q}}

\newcommand{\mathset}[1]{\left\{#1\right\}}
\newcommand{\abs}[1]{\left|#1\right|}
\newcommand{\ceilenv}[1]{\left\lceil #1 \right\rceil}
\newcommand{\floorenv}[1]{\left\lfloor #1 \right\rfloor}
\newcommand{\parenv}[1]{\left( #1 \right)}

\newcommand{\be}[1]{\begin{equation}\label{#1}}
\newcommand{\ee}{\end{equation}}

\renewcommand{\le}{\leqslant}
\renewcommand{\leq}{\leqslant}
\renewcommand{\ge}{\geqslant}
\renewcommand{\geq}{\geqslant}

\newcommand{\Cref}[1]{Co\-ro\-lla\-ry\,\ref{#1}}

\newcommand{\dout}{\Delta}

\DeclareMathOperator{\E}{E}
\DeclareMathOperator{\pr}{Pr}

\outer\def\proclaim #1. #2\par{\medbreak
 \noindent{\bf#1.\enspace}{\sl#2\par}%
 \ifdim\lastskip<\medskipamount \removelastskip\penalty55\medskip\fi}

\begin{document}

\title{\textbf{Trajectory Codes for Flash Memory}}

\author{\large
Anxiao~(Andrew)~Jiang,~\IEEEmembership{Member,~IEEE,}
Michael~Langberg,~\IEEEmembership{Member,~IEEE,}
Moshe~Schwartz,~\IEEEmembership{Senior~Member,~IEEE,} and Jehoshua
Bruck,~\IEEEmembership{Fellow,~IEEE}
\hspace{-5.5ex}
\thanks{The material in this paper was presented in part at the
IEEE International Symposium on Information
Theory (ISIT 2009), Seoul, South Korea, June 2009.}
\thanks{Anxiao (Andrew) Jiang is with the Department of Computer Science and Engineering,
  Texas A\&M University, College Station, TX 77843-3112, U.S.A.
  (e-mail: ajiang@cse.tamu.edu).}
\thanks{Michael Langberg is with the Computer Science Division, Open University of Israel, Raanana 43107, Israel
  (e-mail: mikel@openu.ac.il).}
\thanks{Moshe Schwartz is with the Department
   of Electrical and Computer Engineering, Ben-Gurion University,
   Beer Sheva 84105, Israel
   (e-mail: schwartz@ee.bgu.ac.il).}
\thanks{Jehoshua Bruck is with the Department of Electrical Engineering,
   California Institute of Technology,
   1200 E. California Blvd.,
   Mail Code 136-93, Pasadena, CA 91125, U.S.A.
   (e-mail: bruck@paradise.caltech.edu).}
\thanks{This work was supported in part by the NSF CAREER Award
CCF-0747415, the NSF grant ECCS-0802107, the ISF grant 480/08, the Open University of Israel Research Fund (grants no. 46109 and 101163), the
GIF grant 2179-1785.10/2007, and the Caltech Lee Center for
Advanced Networking.}
}

\maketitle

\begin{abstract}
Flash memory is well-known for its inherent asymmetry:
the flash-cell charge levels are easy to increase but are hard to
decrease. In a general rewriting model, the stored data changes
its value with certain patterns. The patterns of data updates are
determined by the data structure and the application, and are
independent of the constraints imposed by the storage medium. Thus,
an appropriate coding scheme is needed so
that the data changes can be updated and stored efficiently under
the storage-medium's constraints.

In this paper, we define the general rewriting problem using a
graph model. It extends many known rewriting models such as
floating codes, WOM codes, buffer codes, etc. We present a new
rewriting scheme for flash memories, called the \emph{trajectory code},
for rewriting the stored data as many times as possible without
block erasures. We prove that the trajectory code is
asymptotically optimal in a wide range of scenarios.

We also present randomized rewriting codes optimized for expected
performance (given \emph{arbitrary}
rewriting sequences).
Our rewriting codes are shown to be
asymptotically optimal.
\end{abstract}

\begin{IEEEkeywords}
flash memory, asymmetric memory, rewriting,
write-once memory, floating codes, buffer codes
\end{IEEEkeywords}

\section{Introduction}\label{section:Introduction}

\IEEEPARstart{M}{any} storage media have constraints on their
state transitions. A typical example is flash memory, the most
widely-used type of non-volatile electronic
memory~\cite{CapGolOliZan99}. A flash memory consists of
floating-gate cells, where a cell uses the charge it stores to
represent data. The amount of charge stored in a cell can be
quantized into $q \ge 2$ discrete values in order to represent up
to $\log_{2}q$ bits. (The cell is called a \emph{single-level cell
(SLC)} if $q=2$, and called a \emph{multi-level cell (MLC)} if $q
> 2$). We call the $q$ states of a cell its \emph{levels}: level
$0$, level $1$, \dots, level $q-1$. The level of a cell can be
increased by injecting charge into the cell, and decreased by
removing charge from the cell. Flash memories have the prominent
property that although it is relatively easy to increase a cell's
level, it is very costly to decrease it. This follows from the
fact that flash-memory cells are organized as blocks, where every
block has about $10^{5}\sim 10^{6}$ cells. To decrease any cell's
level, the whole block needs to be erased (which means to remove
the charge from all the cells of the block) and then be
reprogrammed. Block erasures not only are slow and energy
consuming, but also significantly reduce the longevity of flash
memories, because every block can endure only $10^{4}\sim 10^{5}$
erasures with guaranteed quality~\cite{CapGolOliZan99}. Therefore,
it is highly desirable to minimize the number of block erasures.
In addition to flash memories, other storage media often have
their own distinct constraints for state transitions. Examples
include magnetic recording~\cite{KurtasBook}, optical
recording~\cite{PohlmannBook}, and phase-change
memories~\cite{RaouxBook}.

In general, the constraints of a memory on its state transitions
can be described by a directed graph, where the vertices represent
the memory states and the directed edges represent the feasible
state transitions~\cite{FiaSha84,FuVin99}. Different edges may
have different costs~\cite{FuYeu00}. Based on the constraints, an
appropriate coding scheme is needed to represent the data so that
the data can be rewritten efficiently. In this paper, we focus on
flash memories, and our objective is to rewrite data as many times
as possible between two block erasures. Note that between two
block erasures, the cell levels can only increase. Therefore we
use the following flash-memory model:

\begin{definition}\label{definition:4}
(\textsc{Flash-Memory Model})

\noindent Consider $n$ flash-memory cells of $q$ levels. The
cells' state can be described by a vector
\[(c_{1},c_{2},\dots,c_{n})\in \{0,1,\dots,q-1\}^{n},\] where for
$i=1,2,\dots,n$, $c_{i}$ is the level of the $i$-th cell. The
cells can transit from one state $(c_{1},c_{2},\dots,c_{n})$ to
another state $(c_{1}',c_{2}',\dots,c_{n}')$ if and only if for
$i=1,2,\dots,n$, $c_{i}'\ge c_{i}$. (If $c_{i}'\ge c_{i}$ for
$i=1,2,\dots,n$, we say that $(c_{1}',c_{2}',\dots,c_{n}')$ is
above $(c_{1},c_{2},\dots,c_{n})$.) \hfill $\Box$
\end{definition}

In this work, we focus on designing rewriting codes for
\emph{general} data-storage applications. How the stored data can
change its value with each rewrite, which we call the
\emph{rewriting model}, depends on the data-storage application
and the used data structure. Several more specific rewriting
models have been studied in the past, including \emph{write-once
memory (WOM)
codes}~\cite{CohGodMer86,FiaSha84,FuVin99,Mer84,RivSha82,WolWynZivKor84},
\emph{floating
codes}~\cite{FinLiuMit08,JiaBohBru07,JiaBru08,MahSieVarWolYaa09,YaaVarSieWol08}
and \emph{buffer codes}~\cite{BohJiaBru07,YaaSieWol08}. In WOM
codes, with each rewrite, the data can change from any value to
any other value. In floating codes, $k$ variables
$v_{1},v_{2},\dots,v_{k}$ are stored, and every rewrite can change
only one variable's value. The rewriting model of floating codes
can be used in many applications where different data items can be
updated individually, such as the data in the tables of databases,
in variable sets of programs, in repeatedly edited files, etc. In
buffer codes, $k$ data items are stored in a queue (namely,
first-in-first-out), and every rewrite inserts a new data item
into the queue and removes the oldest data item.

All the above rewriting models can be generalized with the
following graph model, which we call the \emph{generalized
rewriting model}.

\begin{definition}\label{definition:5}
(\textsc{Generalized Rewriting Model})

\noindent The stored data and the possible rewrites are
represented by a directed graph
\[\cD=(V_{\cD},E_{\cD}).\] The vertices $V_{\cD}$ represent all the
values that the data can take. There is a directed edge $(u,v)$
from $u\in V_{\cD}$ to $v\in V_{\cD}$ (where $v\neq u$) iff a
rewrite may change the stored data from value $u$ to value $v$.
The graph $\cD$ is called the \emph{data graph}, and its number of
vertices -- which corresponds to the data's alphabet size -- is
denoted by \[L=\abs{V_{\cD}}.\] (Throughout the paper we assume
that the data graph is strongly connected.) \hfill $\Box$
\end{definition}

\begin{figure*}[t]
\centering
\includegraphics[width=1.8\columnwidth]{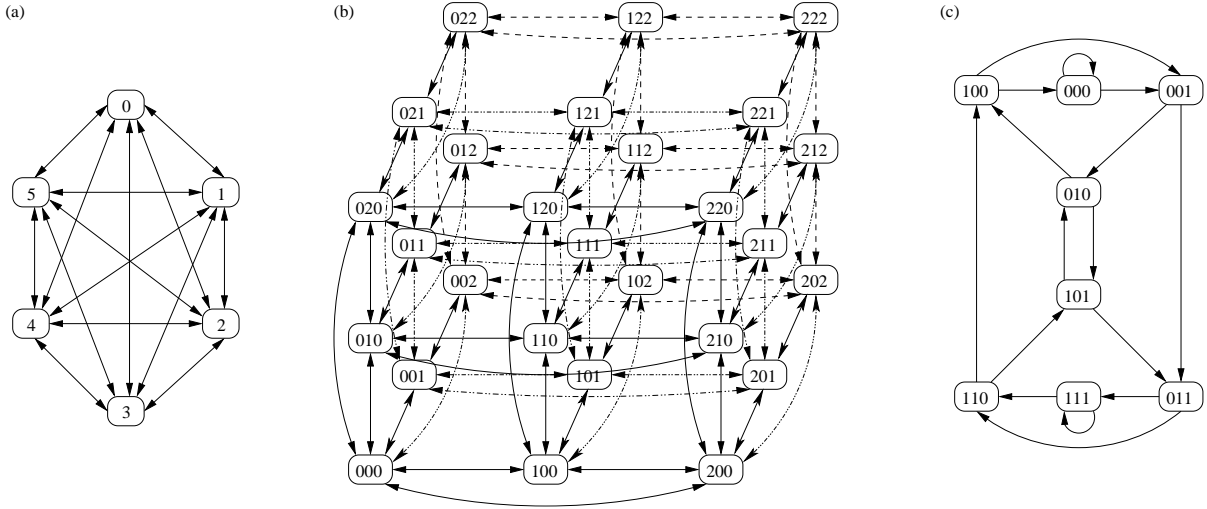}
\caption{The data graph $\cD$ for different rewriting models. (a)
The data graph $\cD$ for a WOM code. Here the data has an alphabet
of size $6$. Since a rewrite can change the data
from any value to any other value, $\cD$ is a complete graph. (b)
The data graph $\cD$ for a floating code. Here $k=3$ variables of
alphabet size $\ell=3$ are stored. Since every rewrite can change
exactly one variable's value, $\cD$ is a generalized hypercube of
regular degree $k(\ell-1)=6$ (for both out-degree and in-degree)
in $k=3$ dimensions. (c) The data graph $\cD$ for a buffer code.
Here $k=3$ variables of alphabet size $\ell=2$ are stored in a
queue. Since every rewrite inserts a new variable into the queue
and removes the oldest variable from the queue, $\cD$ is a de
Bruijn graph of degree $\ell=2$.} \label{fig:1} \vspace{-.1in}
\end{figure*}

Note that the data graph is a complete graph for WOM codes, a
generalized hypercube for floating codes, and a de Bruijn graph
for buffer codes. Some examples are shown in Fig.~\ref{fig:1}.
With more data storage applications and data structures, the data
graph can vary even further. This motivates us to study rewriting
codes for the generalized rewriting model.

A rewriting code for flash memories can be formally defined as
follows. Note that in the flash-memory model, $n$ cells of $q$
levels are used. The definition below can be easily extended to
other constrained memory models.

\begin{definition}\label{definition:3}
(\textsc{Rewriting Code})

\noindent A rewriting code has a \emph{decoding function} $F_{d}$
and an \emph{update function} $F_{u}$. The decoding function
\[F_{d}: \{0,1,\dots,q-1\}^{n}\to V_{\cD}\] means that the cell
state $s\in \{0,1,\dots,q-1\}^{n}$ represents the data
$F_{d}(s)\in V_{\cD}$. The update function (which represents a
rewrite operation), \[F_{u}: \{0,1,\dots,q-1\}^{n} \times V_{\cD}
\to \{0,1,\dots,q-1\}^{n},\] means that if the current cell state
is $s\in \{0,1,\dots,q-1\}^{n}$ and the rewrite changes the data
to $v\in V_{\cD}$, then the rewriting code changes the cell state
to $F_{u}(s,v)$. All the following must hold:
\begin{enumerate}
\item
$(F_{d}(s),v)\in E_{\cD}$.
\item
The cell-state vector
$F_{u}(s,v)$ is above $s$.
\item
$F_{d}(F_{u}(s,v))=v$.
\end{enumerate}
Note that if $F_{d}(s)=v$, we may set $F_{u}(s,v)=s$, which
corresponds to the case where we do not need to change the stored
data. Throughout the paper we do not consider such a case as a
rewrite operation. \hfill $\Box$
\end{definition}

A sequence of rewrites is a sequence $(v_{0},v_{1},v_{2}\dots)$
such that the $i$-th rewrite changes the stored data from
$v_{i-1}$ to $v_{i}$. Given a rewriting code $\cC$, we denote by
$t(\cC)$ the maximal number of rewrites that $\cC$ guarantees to support
for all rewrite sequences. Thus, $t(\cC)$ is a worst-case
performance measure of the code. The code $\cC$ is said to be
\emph{optimal} if $t(\cC)$ is maximized.
In addition to this
definition, if a probabilistic model for rewrite sequences is considered, the expected rewriting
performance can be defined accordingly.

In this paper, we study generalized rewriting for the flash-memory
model. We present a novel rewriting code, called the
\emph{trajectory code}, which is provably asymptotically optimal
(up to constant factors) for a very wide range of scenarios. The
idea of the code is to trace the changes of data in the data graph
$\mathcal{D}$. The trajectory code includes WOM codes, floating
codes, and buffer codes as special cases.

We also study randomized rewriting codes and design codes that are optimized for the
expected rewriting performance (namely, the expected number of
rewrites the code supports). A rewriting code is called
\emph{robust} if its expected rewriting performance is
asymptotically optimal for \emph{all} rewrite sequences.
We present a randomized code construction that is
 robust.

Both our codes for general rewriting and our robust code are optimal up to constant factors (factors independent of the problem parameters).
Namely, for a constant $r \leq 1$, we present codes $\cC$ for which $t(\cC)$ is at least $r$ times that of the optimal code.
We would like to note that, for our robust code,
the constant involved is arbitrarily close to $1$.

The rest of the paper is organized as follows. In
Section~\ref{section:Overview}, we review the related results on
rewriting codes, and compare them to the results derived in this
paper. In Section~\ref{section:trajectoryCode}, a new rewriting
code for the generalized rewriting model, the \emph{trajectory
code}, is presented and its optimality is proved.
In Section~\ref{section:robustCode}, robust codes
optimized for expected rewriting performance are presented. In
Section~\ref{section:conclusions}, the concluding remarks are
presented.

\section{Overview of Related Results}\label{section:Overview}

There has been a history of distinguished theoretical study on
constrained memories. It includes the original work by Kuznetsov
and Tsybakov on coding for defective memories~\cite{KuzTsy74}.
Further developments on defective memories
include~\cite{HeeGam83,KuzVin94}.  The write-once memory
(WOM)~\cite{RivSha82}, write-unidirectional memory
(WUM)~\cite{Ove91,Sim89,WilVinck86}, and write-efficient
memory~\cite{AhlZha94,FuYeu00}, are also special instances of
constrained memories. Among them, WOM is the most related to the
flash-memory model studied in this paper.

Write-once memory (WOM) was studied by Rivest and Shamir in their
original work~\cite{RivSha82}. In a WOM, a cell's state can change
from 0 to 1 but not from 1 to 0. This model was later generalized
with more cell states in~\cite{FiaSha84,FuVin99}. The objective of
WOM codes is to maximize the number of times that the stored data
can be rewritten. A number of very interesting WOM code
constructions have been presented over the years, including the
tabular codes, linear codes, and others in~\cite{RivSha82}, the
linear codes in~\cite{FiaSha84}, the codes constructed using
projective geometries~\cite{Mer84}, and the coset coding
in~\cite{CohGodMer86}. Profound results on the capacity of WOM
have been presented
in~\cite{FuVin99,Hee85,RivSha82,WolWynZivKor84}. Furthermore,
error-correcting WOM codes have been studied in~\cite{ZemCoh91}.
In all the above works, the rewriting model assumes no constraints
on the data, namely, the data graph $\cD$ is a complete graph.

With the increasing importance of flash memories, the flash-memory
model was proposed and studied recently
in~\cite{BohJiaBru07,JiangEccWomISIT07,JiaBohBru07}. The rewriting
schemes include floating
codes~\cite{JiangEccWomISIT07,JiaBohBru07,FloatingBufferJournal,JiaBru08}
and buffer codes~\cite{BohJiaBru07,FloatingBufferJournal}. Both
types of codes use the joint coding of multiple variables for
better rewriting capability. Their data graphs $\cD$ are
generalized hypercubes and de Bruijn graphs, respectively.
Multiple floating codes have been presented, including the code
constructions in~\cite{JiaBohBru07,JiaBru08}, the flash codes
in~\cite{MahSieVarWolYaa09,YaaVarSieWol08}, and the constructions
based on Gray codes in~\cite{FinLiuMit08}. The floating codes
in~\cite{FinLiuMit08} were optimized for the expected rewriting
performance. The study of WOM codes -- with new applications to
flash memories -- is also continued, with a number of improved
code
constructions~\cite{KayserAllerton10,WuISIT10,WuJiangPositionModulation,YaakobiTwoWriteWOMITW10,YaakobiMultipleWOMECC}.

Compared to existing codes, the codes in this paper not only work
for a more general rewriting model, but also provide
efficiently encodable and decodable asymptotically-optimal performance for a wider range of cases.
This can be seen clearly from Table~\ref{tab:1}, where the
asymptotically-optimal codes are summarized. We explain some of
the parameters in Table~\ref{tab:1} here. For the WOM code, a
variable of alphabet size $\ell$ is stored. For the floating code
and the buffer code, $k$ variables of alphabet size $\ell$ are
stored. For rewriting codes using the generalized rewriting model,
$L$ is the size of the data graph. For all the codes, $n$ cells
are used to store the data. It can be seen that this paper
substantially expands the known results on rewriting codes.

\begin{table*}
\caption{A summary of the rewriting codes with asymptotically
optimal performance (up to constant factors). Here $n,k,\ell,L$
are as defined in Section~\ref{section:Introduction} and
Section~\ref{section:Overview}.}
\label{tab:1}
\begin{center}
\begin{small}
\begin{tabular}{|l|l|l|}
\hline 
 \textsc{Type} & \textsc{Asymptotic optimality} & \textsc{ref.}\\
\hline \hline
WOM code ($\cD$ is a complete graph) & $t(\cC)$ is asymptotically optimal  & \cite{RivSha82}\\
\hline
WOM code ($\cD$ is a complete graph) & $t(\cC)$ is asymptotically optimal when $\ell=\Theta(1)$ & \cite{FiaSha84} \\
\hline
Floating code ($\cD$ is a hypercube) & $t(\cC)$ is asymptotically optimal when $k=\Theta(1)$ and $\ell=\Theta(1)$  & \cite{JiaBohBru07} \cite{JiaBru08} \\
\hline Floating code ($\cD$ is a hypercube) & $t(\cC)$ is
asymptotically optimal when $n=\Omega(k\log k)$ and
$\ell=\Theta(1)$ & \cite{JiaBohBru07} \cite{JiaBru08} \\ \hline
Floating code ($\cD$ is a hypercube) & $t(\cC)$ is asymptotically optimal when $n=\Omega(k^{2})$ and $\ell=\Theta(1)$ & \cite{YaaVarSieWol08} \\
\hline
Buffer code ($\cD$ is a de Bruijn graph)  & $t(\cC)$ is asymptotically optimal when $n=\Omega(k)$ and $\ell=\Theta(1)$ & \cite{BohJiaBru07} \cite{YaaSieWol08} \\
\hline
Floating code ($\cD$ is a hypercube) & codes designed for random rewriting sequences when $k=\Theta(1)$ and $\ell=2$ & \cite{FinLiuMit08} \\
\hline WOM code ($\cD$ is a complete graph)     & $t(\cC)$ is asymptotically optimal & this paper
\\
\hline
Rewriting code for the generalized  & For any $\Delta$, $t(\cC)$ is asymptotically optimal when $n=\Omega(\log L)$ & this paper       \\
rewriting model ($\cD$ has maximum out-                    & &  \\
degree $\dout$.)                           &     &          \\
\hline Robust coding  & Asymptotically optimal (with constant $1-\varepsilon$) when $nq = \Omega(L\log L)$
  & this paper \\
             \hline
\end{tabular}
\end{small}
\end{center}
\end{table*}

\section{Trajectory Code}\label{section:trajectoryCode}

We use the flash-memory model of Definition~\ref{definition:4} and
the generalized rewriting model of Definition~\ref{definition:5}
in the rest of this paper. We first present a novel code
construction, the \emph{trajectory code}, then show its performance is
asymptotically optimal.

\subsection{Trajectory Code Outline}
\label{sec:outline}

Let $n_{0},n_{1},n_{2},\dots,n_{d}$ be $d+1$ positive integers and
let \[n=\sum_{i=0}^{d}n_{i},\] where $n$ denotes the number of
flash-memory cells, each of $q$ levels. We partition the $n$ cells
into $d+1$ groups, each with $n_{0}, n_{1},\dots,n_{d}$ cells,
respectively. We call them \emph{registers}
$S_{0},S_{1},\dots,S_{d}$, respectively.

Our encoding uses the following basic scheme: we start by using
register $S_{0}$, called the \emph{anchor}, to record the value of
the initial data $v_{0}\in V_{\cD}$.

For the next $d$ rewrite
operations we use a differential scheme: denote by
$v_1,\dots,v_d\in V_{\cD}$ the next $d$ values of the rewritten
data. In the $i$-th rewrite, $1\leq i\leq d$, we store in register
$S_i$ the identity of the edge $(v_{i-1},v_i)\in E_{\cD}$. We do
not require a unique label for all edges globally, but rather
require that \emph{locally}, for each vertex in $V_{\cD}$, its
out-going edges have unique labels from $\mathset{1,\dots,\dout}$,
where $\dout$ denotes the maximal out-degree in the data graph
$\cD$.

Intuitively, the first $d$ rewrite operations are achieved by
encoding the \emph{trajectory} taken by the input sequence
starting with the anchor data. After $d$ such rewrites, we repeat
the process by rewriting the next input from $V_{\cD}$ in the
anchor $S_0$, and then continuing with $d$ edge labels in
$S_1,\dots,S_d$.

Let us assume a sequence of $s$ rewrites have been stored thus
far.  To decode the last stored value, all we need to know is
$s \mod (d+1)$.
This is easily achieved by using $\ceilenv{t/q}$
more cells (not specified in the previous $d+1$ registers), where
$t$ is the total number of rewrite operations we would like to
guarantee. For these $\ceilenv{t/q}$ cells we employ a simple
encoding scheme: in every rewrite operation we arbitrarily choose
one of those cells and raise its level by one. Thus, the total
level in these cells equals $s$.

The decoding process takes the value of the anchor $S_0$ and then
follows $(s-1)\mod(d+1)$ edges which are read consecutively from
$S_1,S_2,\dots$. Notice that this scheme is appealing in cases
where the maximum out-degree of $\cD$ is significantly lower than
the size of the state space $|V_{\cD}|$.

Note that for $i=0,1,\dots,d$, each register $S_{i}$ can be seen
as a \emph{smaller rewriting code} whose data graph is a
\emph{complete graph} of either $L$ vertices (for $S_{0}$) or
$\dout$ vertices (for $S_{1},\dots,S_{d}$). We let $d=0$ if $\cD$
is a complete graph, and describe how to set $d$ when $\cD$ is not
a complete graph in section~\ref{sec:bounded}. The encoding used
by each register is described in the next section.

\subsection{Analysis for a Complete Data Graph}
\label{sec:clique}

In this section we present an efficiently encodable and decodable
code that enables us to store and rewrite symbols from an input
alphabet $V_{\cD}$ of size $L\geq 2$, when $\cD$ is a
complete graph. The information is stored in $n$ flash-memory
cells of $q$ levels each.

We first state a scheme that allows approximately $nq/8$ rewrites in
the case in which $2\leq L \leq n$. We then extend it to hold for
general $L$ and $n$. We present the quality of our code constructions
(namely the number of possible rewrites they perform) using asymptotic
notation: $O(\cdot)$, $\Omega(\cdot)$, $\Theta(\cdot)$, $o(\cdot)$,
and $\omega(\cdot)$ (where in all cases $n$ is considered to be the
asymptotic variable that tends to infinity).

\subsubsection{The case of $2\leq L \leq n$}
\label{sec:linear} In this section we present a code for small
values of $L$. The code we present is essentially the one
presented in \cite{RivSha82}.

\begin{construction}\label{construction:4}
Let $2\leq L \leq n$.  This construction produces an efficiently
encodable and decodable rewriting code $\cC$ for a complete data
graph $\cD$ with $L$ states, and flash memory with $n$ cells with
$q$ levels each.

Let us first assume $n=L$. Denote the $n$ cell levels by
$\vec{c}=(c_{0},c_{1},\dots,c_{L-1})$, where $c_{i}\in
\{0,1,\dots,q-1\}$ is the level of the $i$-th cell for
$i=0,1,\dots,L-1$. Denote the alphabet of the data by
$V_{\cD}=\{0,1,\dots,L-1\}$. We first use only cell levels 0 and
1, and the data stored in the cells is
\[\sum_{i=0}^{L-1}ic_{i}\pmod{L}.\] With each rewrite, we increase
the minimum number of cell levels from 0 to 1 so that the new cell
state represents the new data. (Clearly, $c_{0}$ remains untouched
as 0.) When the code can no longer support rewriting, we increase
all cells (including $c_{0}$) from 0 to 1, and start using cell
levels 1 and 2 to store data in the same way as above, except that
the data stored in the cells uses the formula
\[\sum_{i=0}^{L-1}i(c_{i}-1)\pmod{L}.\] This process is repeated
$q-1$ times in total. The general decoding function is therefore
defined as
\[F_d(\vec{c})=\sum_{i=0}^{L-1}i(c_{i}-c_{0})\pmod{L}.\]

We now extend the above code to $n\ge L$ cells. We divide the $n$
cells into $b=\floorenv{n/L}$ groups of size $L$ (some cells may
remain unused). We first apply the code above to the first group
of $L$ cells, then to the second group, and so on. \hfill $\Box$
\end{construction}

\begin{theorem}
\label{the:small} Let $2\leq L \leq n$. The number of rewrites
the code $\cC$ of
Construction~\ref{construction:4} guarantees is lower bounded by
\[t(\cC)\geq n(q-1)/8=\Omega(nq).\]
\end{theorem}
\begin{IEEEproof}
First assume $n=L$. When cell levels $j-1$ and $j$ are used to
store data (for $j=1,\dots,q-1$), by the analysis
in~\cite{RivSha82}, even if only one or two cells increase their
levels with each rewrite, at least $(L+4)/4$ rewrites can be
supported. So the $L$ cells can support at least
\[t(\cC)\geq \frac{(L+4)(q-1)}{4}=\Omega(nq)\]
rewrites. Now let $n\ge L$. When
$b=\floorenv{n/L}$, it is easy to see that $bL\geq n/2$. The $b$
groups of cells can guarantee
\[t(\cC)\geq\frac{b(L+4)(q-1)}{4} \geq \frac{n(q-1)}{8} = \Omega(nq)\]
rewrites.
\end{IEEEproof}

\subsubsection{The Case of Large $L$}
\label{sec:general} We now consider the setting in which $L$ is
larger than $n$.  The rewriting code we present reduces the
general case to that of the case $n=L$ studied above.  The
majority of our analysis addresses the case in which $n < L \leq
2^{n/16}$.  We start, however, by first considering the simple
case in which $2^{n/16} \leq L \leq q^n$. Notice that if $L$ is
greater than $q^n$ then we cannot guarantee even a single rewrite.

\begin{construction}\label{construction:trivial}
Let $c \in [2^{1/16},q]$, and let $L = c^n$.
This construction produces an efficiently
encodable and decodable rewriting code $\cC$ for a complete data
graph $\cD$ with $L$ states, and flash memory with $n$ cells with
$q$ levels each.

Denote the $n$ cell levels by
$\vec{c}=(c_{0},c_{1},\dots,c_{n-1})$, where $c_{i}\in
\{0,1,\dots,q-1\}$ is the level of the $i$-th cell for
$i=0,1,\dots,n-1$. Denote the alphabet of data by
$V_{\cD}=\{0,1,\dots,L-1\}$. For the initial (re)write we use only
cell levels 0 to $\ceilenv{c}-1$, and the data stored in the
cells is \[\sum_{i=0}^{n-1}c_{i}\ceilenv{c}^i\pmod{L}.\] With
the next rewrite, we use the cell levels $\ceilenv{c}$ to
$2\ceilenv{c}-1$ and the data stored in the cells is now
\[\sum_{i=0}^{n-1}(c_{i}-\ceilenv{c})\ceilenv{c}^i\pmod{L}\]
and so on. In general,
\[F_d(\vec{c})=\sum_{i=0}^{n-1}(c_{i}\bmod \ceilenv{c})\ceilenv{c}^i\pmod{L}.\]
and with each rewrite we represent $v \in V_{\cD}$ by its $n$-character representation over an alphabet of size
$\ceilenv{c}$.\hfill $\Box$
\end{construction}

The following theorem is immediate.

\begin{theorem}
Let $c \in [2^{1/16},q]$.
If $L = c^n$ then the code $\cC$ of Construction~\ref{construction:trivial} guarantees
$t(\cC)\geq q/\ceilenv{c} = \Omega(q/c)$.
\end{theorem}

We now address the case $n < L \leq 2^{n/16}$.
Let $b$ be the smallest positive integer value that satisfies
\[\floorenv{n/b}^b \geq L.\]

\begin{claim}
\label{claim:b1} For $n \leq L \leq 2^{n/16}$, it holds
that
\[b \leq \frac{2\log{L}}{\log{(n/\log{L})}}.\]
\end{claim}

\begin{IEEEproof}
Let $b = \frac{2\log{L}}{\log{(n/\log{L})}}$.
Notice that
\[
\floorenv{n/b} \geq \frac{n \log{(n/\log{L})}}{4\log{L}}.
\]
Thus,
\begin{align*}
\log{\floorenv{n/b}^b}
& =
b\log{\floorenv{n/b}} \\
& \geq
\frac{2\log{L}}{\log{(n/\log{L})}}\log\parenv{\frac{n \log{(n/\log{L})}}{4\log{L}}} \\
& \geq
\frac{2\log{L}}{\log{(n/\log{L})}}\log\parenv{\frac{n}{4\log{L}}} \\
& \geq
\frac{2\log{L}}{\log{(n/\log{L})}}\log\parenv{\sqrt{\frac{n}{\log{L}}}} = \log{L}
\end{align*}
We used the fact that $L \leq 2^{n/16}$ to establish the inequality $\frac{n}{4\log{L}} \geq \sqrt{\frac{n}{\log{L}}}$ used in the last step above.
\end{IEEEproof}

\begin{construction}\label{construction:5}
Let $n < L \leq 2^{n/16}$.  This construction produces an
efficiently encodable and decodable rewriting code $\cC$ for a
complete data graph $\cD$ with $L$ states, and flash memory with
$n$ cells with $q$ levels each.

For $i=1,2,\dots,b$, let $v_{i}$ be a symbol from an alphabet of
size \[\floorenv{n/b}\geq L^{1/b}.\] We may represent any symbol
$v\in V_{\cD}$ as a vector of symbols
$(v_{1},v_{2},\dots,v_{b})$.

Partition the $n$ flash-memory cells into $b$ groups, each with
$\floorenv{n/b}$ cells (some cells may remain unused). Encoding
the symbol $v$ into $n$ cells is equivalent to the encoding of
each $v_i$ into the corresponding group of $\floorenv{n/b}$ cells.
As the alphabet size of each $v_i$ equals the number of cells it
is to be encoded into, we can use
Construction~\ref{construction:4} to store $v_{i}$. \hfill $\Box$
\end{construction}

\begin{example}
Let $n=16$, $q=4$, $L=56$, and the data graph $\cD$ be a complete
graph. We design a rewriting code for these parameters with the
method of Construction~\ref{construction:5}.

Let $b=2$, and we divide the $n=16$ cells evenly into $b=2$
groups. Let $\vec{c}=(c_{0},c_{1},\dots,c_{7})$ denote the cell
levels of the first cell group, and let
$\vec{c}'=(c_{0}',c_{1}',\dots,c_{7}')$ denote the cell levels of
the second cell group.

Let $v\in \{0,1,\dots,L-1\}=\{0,1,\dots,55\}$ denote the value of
the stored data. Let $v_{1}$ and $v_{2}$ be two symbols of
alphabet size $8$. We can represent $v$ by the pair $(v_1,v_2)$
as follows:
\[v_{1}=\floorenv{v/8} \qquad\qquad v_{2}=v\bmod 8.\]

We store $v_{1}$ in the first cell group using the decoding
function \[v_{1}=\sum_{i=0}^{7}i(c_{i}-c_{0})\pmod{8},\] and
store $v_{2}$ in the second cell group using the decoding function
\[v_{2}=\sum_{i=0}^{7}i(c_{i}'-c_{0}')\pmod{8}.\]
Reconstructing $v$ from $(v_1,v_2)$ is done by $v=8v_{1}+v_{2}$.
Thus, if the data, $v$, changes as
\[0\to 23 \to 45 \to 6 \to 27 \to 12,\]
the symbol pair $(v_{1},v_{2})$ will change as \[(0,0)\to
(2,7)\to (5,5) \to (0,6)\to (3,3)\to (1,4),\] and the cell levels
$(\vec{c},\vec{c}')=((c_{0},c_{1},\dots,c_{7}),(c_{0}',c_{1}',\dots,c_{7}'))$
will change as
\begin{eqnarray*}
&((0,0,0,0,0,0,0,0),(0,0,0,0,0,0,0,0))\\
&\downarrow\\
&((0,0,1,0,0,0,0,0),(0,0,0,0,0,0,0,1))\\
&\downarrow\\
&((0,0,1,1,0,0,0,0),(0,0,0,0,0,0,1,1))\\
&\downarrow\\
&((0,0,1,1,1,0,0,1),(0,1,0,0,0,0,1,1))\\
&\downarrow\\
&((0,0,1,1,1,1,1,1),(0,1,0,0,0,1,1,1))\\
&\downarrow\\
&((1,2,1,1,1,1,1,1),(0,1,1,1,1,1,1,1))
\end{eqnarray*}

A careful reader will have observed that the parameters here
actually do not satisfy the condition $n < L \leq 2^{n/16}$.
Indeed, the condition $n < L \leq 2^{n/16}$ is chosen only for
the analysis of the asymptotic performance. The rewriting code of
Construction~\ref{construction:5} can be used for more general
parameter settings.
\end{example}

\begin{theorem}
\label{the:large} Let $n \leq L \leq 2^{n/16}$. The number of rewrites the
code $\cC$ of Construction~\ref{construction:5} guarantees
is lower bounded by
\[t(\cC)\geq \frac{n(q-1)\log{(n/\log{L})}}{16\log{L}}=
\Omega\parenv{\frac{nq\log{(n/\log{L})}}{\log{L}}}.\]
\end{theorem}
\begin{IEEEproof}
Using Construction~\ref{construction:5}, the number of rewrites
possible is bounded by the number of rewrites possible for each of
the $b$ cell groups. By Theorem~\ref{the:small} and
Claim~\ref{claim:b1}, this is at least
\begin{align*}
\floorenv{\frac{n}{b}}\cdot\frac{q-1}{8}
& \geq
\parenv{\frac{n\log{(n/\log{L})}}{2\log L}-1}\frac{q-1}{8} \\
& =
\Omega\parenv{\frac{nq\log{(n/\log{L})}}{\log{L}}}.
\end{align*}
\end{IEEEproof}

\subsection{Analysis for a Bounded-Out-Degree Data Graph}
\label{sec:bounded}

We now return to the outline of the trajectory code from Section
\ref{sec:outline}, and apply it in full detail using the codes
from Section \ref{sec:clique} to the case of data graphs $\cD$
with upper bounded out-degree $\dout$. We refer to such graphs as
$\dout$-restricted. To simplify our presentation, in the theorems
below we will again use the asymptotic notation freely; however,
as opposed to the previous section we will no longer state or make
an attempt to optimize the constants involved in our calculations.
We assume that $n\leq L$, since for $L \leq n$,
Construction~\ref{construction:4} can be used to obtain optimal
codes (up to constant factors). In this section we study the case
$L \leq 2^{n/16}$. We do not address the case of larger $L$, as
its analysis, although based on similar ideas, becomes rather
tedious and overly lengthy.

Using the notation of Section~\ref{sec:outline}, to realize the
trajectory code we need to specify
the sizes $n_i$ and the value of $d$.  We consider two cases: the case
in which $\dout$ is \emph{small} compared to $n$, and the case in
which $\dout$ is \emph{large}.

The following construction is for the case in which $\dout$ is
\emph{small} compared to $n$.

\begin{construction}\label{construction:6}
Let
\[\dout \leq \floorenv{\frac{n\log{(n/\log{L})}}{2\log{L}}}.\]
We build
an efficiently encodable and decodable rewriting code $\cC$ for
any $\dout$-restricted data graph $\cD$ with $L$ vertices and $n$
flash-memory cells of $q$ levels as follows. For the trajectory
code, let
\[d=\floorenv{\log L/\log{(n/\log{L})}} = \Theta(\log
L/\log{(n/\log{L})}).\]
Set the size of the $d+1$ registers to
\[n_0=\floorenv{n/2}\] and \[n_i = \floorenv{n/(2d)} \geq \dout\]
for
$i=1,2,\dots, d$.  (We obviously have $\sum_{i=0}^{d} n_i \leq
n$.)

The update and decoding functions of the trajectory code $\cC$ are
defined as follows. We use the encoding scheme specified in
Construction~\ref{construction:5} to store in the $n_{0}$ cells of
the register $S_{0}$ an ``anchor'' (i.e., a vertex) of $\cD$,
which is a symbol in the alphabet $V_{\cD}=\{0,1,\dots,L-1\}$.

For $i=1,2,\dots,d$, we use the encoding scheme specified in
Construction~\ref{construction:4} to store in the $n_{i}$ cells of
the register $S_{i}$ an ``edge'' of $\cD$, which is a symbol in
the alphabet $\{0,1,\dots,\dout -1\}$. Notice that the latter is
possible because $n_i \geq \dout$ for $i=1,\dots d$. \hfill $\Box$
\end{construction}

Recall that
the anchor and the edges stored in $S_{0},S_{1},S_{2},\dots$ show
how the data changes its value with rewrites. That is, they show
the trace of the changing data in the data graph $\cD$. Every
$d+1$ rewrites change the data stored in the register $S_{i}$
exactly once, for $i=0,1,\dots,d$. After every $d+1$ rewrites, the
next rewrite resets the anchor's value in $S_{0}$, and the same
rewriting process starts again.

Suppose that the rewrites change the stored data as $v_{0}\to
\cdots \to v_{i}\to v_{i+1}\to \cdots$. Then with the rewriting
code of Construction~\ref{construction:6}, the data stored in the
register $S_{0}$ changes as $v_{0}\to v_{d+1}\to v_{2(d+1)}\to
v_{3(d+1)}\to \cdots$. For $i=1,2,\dots,d$, the data stored in the
register $S_{i}$ changes as $(v_{i-1},v_{i})\to
(v_{i-1+(d+1)},v_{i+(d+1)})\to (v_{i-1+2(d+1)},v_{i+2(d+1)})\to
(v_{i-1+3(d+1)},v_{i+3(d+1)})\to \cdots$ Here every edge
$(v_{j-1},v_{j})\in E_{\cD}$ is locally labeled by the alphabet
$\{0,1,\dots,\dout-1\}$.

\begin{theorem}
\label{the:s_delta} Let $L \leq 2^{n/16}$ and
$\dout \leq
\floorenv{\frac{n\log{(n/\log{L})}}{2\log{L}}}$. The number
of rewrites the code $\cC$ of
Construction~\ref{construction:6} guarantees is
\[t(\cC)=\Omega(nq)\]
\end{theorem}
\begin{IEEEproof}
By Theorems~\ref{the:large} and \ref{the:small}, the lower bound
on the number of
rewrites possible in $S_0$ is equal (up to constant factors) to that
of $S_i$ ($i \geq 1$):
\begin{align*}
\Omega\parenv{\frac{n_0 q\log{(n_0/\log{L})}}{\log{L}}} & =  \Omega\parenv{\frac{nq\log{(n/\log{L})}}{\log{L}}} \\
& = \Omega\parenv{\frac{nq}{d}} = \Omega\parenv{n_i q}.
\end{align*}
Thus, the total number of rewrites in the scheme outlined in
Section~\ref{sec:outline} is lower bounded by
$d+1$ times the bound for each register
$S_i$, and so $t(\cC)=\Omega(nq)$.
\end{IEEEproof}

\begin{example}
Consider floating codes, where $k$ variables of alphabet size
$\ell$ are stored in $n$ cells of $q$ levels. When
Construction~\ref{construction:6} is used to build the floating
code, we get $L=\ell^{k}$ and $\dout = k(\ell-1)$. So if
$k(\ell-1) \le \floorenv{\frac{n\log{(n/(k\log{\ell}))}}{2k\log \ell}}$, the
code can guarantee $t(\cC)=\Omega(nq)$ rewrites, which is
asymptotically optimal.
\end{example}

The next construction is for the case in which $\dout$ is
\emph{large} compared to $n$.

\begin{construction}\label{construction:7}
Let $L \leq 2^{n/16}$ and
let \[\floorenv{\frac{n\log{(n/\log{L})}}{2\log{L}}} \leq \dout \leq L-1.\]
We build an efficiently encodable and decodable rewriting code
$\cC$ for any $\dout$-restricted data graph $\cD$ with $L$
vertices and $n$ flash-memory cells of $q$ levels as follows. For
the trajectory code, let \[d=\floorenv{\log L/\log{\dout}} =
\Theta(\log L/\log{\dout}).\] Set the size of the registers to
\[n_0=\floorenv{n/2}\] and \[n_i = \floorenv{n/(2d)}\] for
$i=1,2,\dots,d$.

The update and decoding functions of the trajectory code $\cC$ are
defined as follows: use the encoding scheme specified in
Construction~\ref{construction:5} to store an ``anchor'' in
$S_{0}$ and store an ``edge'' in $S_{i}$, for $i=1,2,\dots,d$.
(The remaining details are the same as
Construction~\ref{construction:6}.) \hfill $\Box$
\end{construction}

\begin{theorem}
\label{the:g_delta}
Let $L \leq 2^{n/16}$.
Let $\floorenv{\frac{n\log{(n/\log{L})}}{2\log{L}}} \leq
\dout \leq L-1$. The number of rewrites the code $\cC$ of
Construction~\ref{construction:7} guarantees is lower bounded by
\[t(\cC)=\Omega\parenv{\frac{nq\log{(n/\log{L})}}{\log{\dout}}}.\]
\end{theorem}

\begin{IEEEproof}
By Theorem~\ref{the:large}, the number of rewrites supported in
$S_0$ is lower bounded by
\[
\Omega\parenv{\frac{n_0 q\log{(n_0/\log{L})}}{\log{L}}} = \Omega\parenv{\frac{nq\log{(n/\log{L})}}{\log{L}}}
\]
Similarly, for $i=1,2,\dots,d$, the number of rewrites supported
in $S_i$ is lower bounded by
\begin{align*}
\Omega\parenv{\frac{n_i q\log{(n_i/\log{\dout})}}{\log{\dout}}} & =
\Omega\parenv{\frac{nq\log{(n/\log{L})}}{d\log{\dout}}} \\
& = \Omega\parenv{\frac{nq\log{(n/\log{L})}}{\log{L}}}.
\end{align*}
Thus, as in Theorem~\ref{the:s_delta}, we conclude that
the total number of rewrites in the scheme outlined in
Section~\ref{sec:outline} is lower bounded by $d+1$ times the bound for each
register $S_i$, and so
$t(\cC)=\Omega\parenv{\frac{nq\log{(n/\log{L})}}{\log{\dout}}}$.
\end{IEEEproof}

\subsection{Optimality of the Code Constructions}

We now prove upper bounds on the number of rewrites in general
rewriting schemes, which match the lower bounds induced by our
code constructions. They show that our code constructions are
asymptotically optimal.

\begin{theorem}
\label{the:small2} Any rewriting code $\cC$ that stores symbols
from some data graph $\cD$ in $n$ flash-memory cells of $q$ levels
supports at most
\[t(\cC)\leq n(q-1)=O(nq)\]
rewrites.
\end{theorem}
\begin{IEEEproof}
The bound is trivial. In the best case, all cells are initialized at level
$0$, and every rewrite increases exactly one cell by exactly one level.
Thus, the total number of rewrites is bounded by $n(q-1)=O(nq)$ as claimed.
\end{IEEEproof}

\begin{corollary}
The codes from Constructions \ref{construction:4} and
\ref{construction:6} are asymptotically optimal.
\end{corollary}

For large values of $L$, we can improve the upper bound. First,
let $r$ denote the largest integer such that
\[\binom{r+n-1}{r}<L-1.\] We need the following technical claim.

\begin{claim}
\label{claim:minraise}
Let $L \leq 2^{n/16}$.
For all $1\leq n < L-1$, the following inequality
\[r\geq c\cdot \frac{\log{L}}{\log (n/\log{L})}\]
holds for a sufficiently small constant $c>0$.
\end{claim}

\begin{IEEEproof}
First, it is easy to see that $r \in [1,n]$.
Now
we may use the well-known bound for
all $v\geq u\geq 1$,
\[\binom{v}{u} < \parenv{\frac{ev}{u}}^u,\]
where $e$ is the base of the natural logarithm.
Let $m=n/r$.
It follows that,
\[\binom{r+n-1}{r}
\leq
\binom{r+n}{r}
\leq
\binom{2n}{r}
\leq
\frac{2^re^rn^r}{r^r}.
\]
Hence,
\[\log\binom{r+n-1}{r}
\leq
r\log\parenv{\frac{2en}{r}} =
\frac{n}{m}\log(2em).
\]
Thus, it suffices to prove that
\[
\frac{n}{m}\log(2em) < \log{(L-1)}.
\]
We conclude via basic computations that if
\[m = c'\cdot\frac{n\log{\parenv{{n/\log{L}}}} }{\log{L}}\]
for a sufficiently large constant $c'>0$, then
\[
\binom{r+n-1}{r} \leq L.
\]
\end{IEEEproof}

\begin{theorem}
\label{the:large2} Let $L \leq 2^{n/16}$. When $n < L-1$, any
rewriting code $\cC$ that stores symbols from the complete data
graph $\cD$ in $n$ flash-memory cells of $q$ levels can guarantee at
most
\[t(\cC)=O\parenv{\frac{nq\log{(n/\log{L})}}{\log{L}}}\]
rewrites.
\end{theorem}
\begin{IEEEproof}
Let us examine some state $s$ of the $n$ flash-memory cells,
currently storing some value $v\in V_{\cD}$, i.e., $F_{d}(s)=v$.
Having no constraint on the data graph, the next symbol we want to
store may be any of the $L-1$ symbols $v'\in V_{\cD}$, where
$v'\neq v$.

If we allow ourselves $r$ operations of increasing a single cell
level of the $n$ flash-memory cells by one (perhaps operating on
the same cell more than once), we may reach at most \[\binom{n+r-1}{r}\]
distinct new states. However, by our choice of $r$, we have
$\binom{n+r-1}{r}< L-1$. So we need at least $r+1$ such operations
to realize a rewrite in the worst case. Since we have a total of
$n$ cells with $q$ levels each, the guaranteed number of rewrite operations
is upper bounded by
\[
t(\cC)\leq \frac{n(q-1)}{r+1} =
O\parenv{\frac{nq\log{(n/\log{L})}}{\log L}}.
\]
\end{IEEEproof}

\begin{corollary}
The code from Construction \ref{construction:5}
is asymptotically optimal.
\end{corollary}

\begin{theorem}
\label{the:large3} Let $2^{n/16} \leq L=c^n \leq q^{n}$. Any
rewriting code $\cC$ that stores symbols from the complete data
graph $\cD$ in $n$ flash-memory cells of $q$ levels can guarantee at
most
\[t(\cC)=O\parenv{q/c}\]
rewrites.
\end{theorem}
\begin{IEEEproof}
We follow the proof of Theorem~\ref{the:large2}. In this case we note that
for $\binom{n+r-1}{r}$ to be at least of size $L=c^n$ we need
$r = \Omega\parenv{nc}$.
The proof follows.
\end{IEEEproof}

\begin{corollary}
The code from Construction \ref{construction:trivial}
is asymptotically optimal.
\end{corollary}

\begin{theorem}
\label{the:g_delta2} Let $L \leq 2^{n/16}$.
Let $\dout >
\floorenv{\frac{n\log{(n/\log{L})}}{2\log{L}}}$.  There exist
$\dout$-restricted data graphs $\cD$ over a vertex set of size
$L$, such that any rewriting code $\cC$ that stores symbols from
the data graph $\cD$ in $n$ flash-memory cells of $q$ levels
can guarantee at most
\[t(\cC)=O\parenv{\frac{nq\log{(n/\log{L})}}{\log{\dout}}}\]
rewrites.
\end{theorem}

\begin{IEEEproof}
We start by showing that $\dout$-restricted graphs $\cD$ with
certain properties do not allow rewriting codes $\cC$ that support
more than
$t(\cC)=O\parenv{\frac{nq\log{(n/\log{L})}}{\log{\dout}}}$
rewrites.  We then show that such graphs indeed exist.  This will
conclude our proof.

Let $\cD$ be a $\dout$-restricted graph whose diameter $d$ is at
most $O\parenv{\frac{\log{L}}{\log{\dout}}}$.  Assuming the
existence of such a graph $\cD$, consider (by contradiction) a
rewriting code $\cC$ for the $\dout$-restricted graph $\cD$ that
allows \[t(\cC)=\omega\parenv{\frac{nq\log{(n/\log{L})}}{\log{\dout}}}\]
rewrites.  We use $\cC$ to construct a rewriting code $\cC'$ for a
new data graph $\cD'$ which has the same vertex set
$V_{\cD'}=V_{\cD}$ but is a complete graph. The code $\cC'$ will
allow \[t(\cC')=\omega\parenv{\frac{nq\log{(n/\log{L})}}{\log{L}}}\]
rewrites, a contradiction to Theorem~\ref{the:large2}. This will
imply that our initial assumption regarding the quality of our
rewriting code $\cC$ is false.

The rewriting code $\cC'$ (defined by the decoding function
$F'_{d}$ and the update function $F'_{u}$) is constructed by
\emph{mimicking} $\cC$ (defined by the decoding function $F_{d}$
and the update function $F_{u}$). We start by setting
$F'_{d}=F_{d}$. Next, let $s$ be some state of the flash cells.
Denote $F_{d}(s)=F'_{d}(s)=v_0\in V_{\cD}$. Consider a rewrite
operation attempting to store a new value $v_1\in V_{\cD}$, where
$v_1\neq v_0$. There exists a path in $\cD$ of length $d'$,
where $d'\leq d$, from $v_0$ to $v_1$, which we denote by
\[v_0,u_1,u_2,\dots,u_{d'-1},v_1.\] We now define
\[F'_{u}(s,v_1)=
F_{u}(F_{u}(\dots F_{u}(F_{u}(s,u_1),u_2)\dots,u_{d'-1}),v_1),\]
which simply states that to encode a new value $v_1$ we follow the
steps taken by the code $\cC$ on a short path from $v_0$ to $v_1$
in the data graph $\cD$.

As $\cC$ guarantees
$t(\cC)=\omega\parenv{\frac{nq\log{(n/\log{L})}}{\log{\dout}}}$ rewrites,
the code for $\cC'$ guarantees at least
\[
t(\cC')=\omega\parenv{\frac{nq\log{(n/\log{L})}}{d\log{\dout}}} = \omega\parenv{\frac{nq\log{(n/\log{L})}}{\log{L}}}
\]
rewrites.  Here we use the fact that $d
=O\parenv{\frac{\log{L}}{\log{\dout}}}$.

What is left is to show the existence of data graphs $\cD$ of
maximum out-degree $\dout$ whose diameter $d$ is at most
$O\parenv{\frac{\log{L}}{\log{\dout}}}$.
To obtain such a graph, one may simply take a rooted bi-directed tree of total degree $\dout$ and corresponding depth $O\parenv{\frac{\log{L}}{\log{\dout}}}$.
\end{IEEEproof}

\begin{corollary}
For $L \leq 2^{n/16}$, the code from Construction \ref{construction:7}
is asymptotically optimal.
\end{corollary}

\section{Robust Rewriting Codes}\label{section:robustCode}

It addition to the worst-case rewriting performance, it is also
interesting to design rewriting codes with good expected
performance.
In this section we consider the use of randomized codes to obtain good expected performance for all
rewrite sequences.

Let $\vec{v}=(v_{1},v_{2},v_{3},\dots,v_{n(q-1)})$ denote a sequence of
rewrites. That is, for $i=1,2,3,\dots,n(q-1)$, the $i$-th rewrite changes
the data to the value $v_{i}\in \{0,1,\dots,L-1\}$. By default,
the original value of the data is $v_{0}=0$, and since every
rewrite changes the data, we require that for all $i\geq 1$,
$v_{i}\ne v_{i-1}$. Also, as no more than $n(q-1)$ rewrites may be supported,
the sequence $\vec{v}$ is limited to $n(q-1)$ elements.

Let $\cC$ denote a rewriting code, which
stores the data from an alphabet of size $L$ in $n$ cells of $q$ levels.
The code $\cC$ can only support a finite number of rewrites in the
rewrite sequence $\vec{v}$. We use $t(\cC|\vec{v})$ to denote
the number of rewrites in the rewrite sequence $\vec{v}$ that are
supported by the code $\cC$. That is, if the code $\cC$ can
support the rewrites $v_{1},v_{2},\dots,$ up to $v_{k}$, then
$t(\cC|\vec{v})=k$.

Let $V$ denote the set of all possible rewrite sequences. If we
are interested in the number of rewrites that a code $\cC$
guarantees in the worst case, $t(\cC)$, then we can
see that \[t(\cC)=\min_{\vec{v}\in V}t(\cC|\vec{v}).\]

In this
section, we are interested in the expected number of rewrites that
a code $\cC$ can support under random coding.
Let $\cQ$ be some distribution over rewriting codes and let $\cC_\cQ$ be a {\em randomized code} (namely, a random variable) with distribution $\cQ$.
Let $\E(x)$ denote the expected value of a random variable
$x$. We define the \emph{expected performance} of the randomized
rewriting code $\cC_\cQ$ to be
\[\E_{\cC_\cQ}=\min_{\vec{v}\in V}\E(t(\cC_\cQ|\vec{v})).\]
Our objective
is to maximize $\E_{\cC_\cQ}$.
Namely, to construct a distribution $\cQ$ such that for all $\vec{v}$, $\cC_\cQ$ will allow many rewrites in expectation.
A code $\cC_\cQ$ whose
$\E_{\cC_\cQ}$ is asymptotically optimal is called a
\emph{robust code}.
For any constant $\varepsilon >0$, in this section we will present a
randomized code with $\E_{\cC_\cQ}\geq(1-\varepsilon)(q-1)n$
(clearly, the code is robust).

\subsection{Code Construction}

We first present our code construction, analyze its
properties and define some useful terms. We then turn to show that it is indeed robust.

Let $(c_{1},c_{1},\dots,c_{n})$ denote the $n$ cell levels, where
for $i=1,2,\dots,n$, $c_{i}\in \{0,1,\dots,q-1\}$ is the $i$-th
cell's level. Given a cell state
$\vec{c}=(c_{1},c_{2},\dots,c_{n})$, we define its \emph{weight},
denoted by $w(\vec{c})$, as \[w(\vec{c})=\sum_{i=1}^{n}c_{i}.\]
Clearly, $0 \le w(\vec{c})\le (q-1)n$. Given the decoding function,
$F_{d}: \{0,1,\dots,q-1\}^{n}\to \{0,1,\dots,L-1\}$,
of a rewriting code, the cell
state $\vec{c}$ represents the data $F_{d}(\vec{c})$.

\begin{construction}\label{constr:01}
For all $i=0,1,\dots,n(q-1)-1$ and $j=1,2,\dots,n$, let
$\theta_{i,j}$ and $a_i$ be parameters chosen from the set
$\{0,1,\dots,L-1\}$.

We define a rewriting code $\cC$ as follows. Its decoding function
is
\[F_{d}(\vec{c})=\parenv{\sum_{i=1}^{n}\theta_{w(\vec{c})-1,i}c_{i}+\sum_{i=0}^{w(\vec{c})-1}a_{i}}
\bmod L.\] By default, if $\vec{c}=(0,0,\dots,0)$, then
$F_{d}(\vec{c})=0$. When rewriting the data, we take a greedy
approach: For every rewrite, minimize the increase of the cell
state's weight. (If there is a tie between cell states of the same
weight, break the tie arbitrarily.) \hfill $\Box$
\end{construction}

For simplicity, we will omit the term ``$\bmod \  L$'' in all
computations below that
consist of values of data. For example, the expression for $F_d$
in the above code construction will be simply written as
\[F_{d}(\vec{c})=\sum_{i=1}^{n}\theta_{w(\vec{c})-1,i}c_{i}+\sum_{i=0}^{w(\vec{c})-1}a_{i},\]
and $F_{d}(\vec{c})- F_{d}(\vec{c'})$ will mean
$(F_{d}(\vec{c})- F_{d}(\vec{c'}))\bmod L$.

\begin{definition}
(\textsc Update Vector and Update Diversity)

\noindent Let $\vec{c}=(c_{1},c_{2},\dots,c_{n})$ be a cell state
where for $i=1,2,\dots,n$, $c_{i}\in \{0,1,\dots,q-2\}$. For
$i=1,2,\dots,n$, we define $N_{i}(\vec{c})$ as
\[N_{i}(\vec{c})=(c_{1},\dots,c_{i-1},c_{i}+1,c_{i+1},\dots,c_{n})\]
and define $e_{i}(\vec{c})$ as
\[e_{i}(\vec{c})=F_{d}(N_{i}(\vec{c}))-F_{d}(\vec{c}).\]
We also define the \emph{update vector} of $\vec{c}$, denoted by $u(\vec{c})$,
as
\[u(\vec{c})=(e_{1}(\vec{c}),e_{2}(\vec{c}),\dots,e_{n}(\vec{c})),\]
and the \emph{update diversity} of $\vec{c}$ as
\[\abs{\mathset{e_{1}(\vec{c}),e_{2}(\vec{c}),\dots,e_{n}(\vec{c})}}.\]
\hfill $\Box$
\end{definition}

The update diversity of a cell state $\vec{c}$ is at most $L$. If
it is $L$, it means that when the current cell state is $\vec{c}$,
no matter what the next rewrite is, we only need to increase one
cell's level by one to realize the rewrite. Specifically, if the
next rewrite changes the data from $F_{d}(\vec{c})$ to $v'$, we
will change from $\vec{c}$ to $N_{i}(\vec{c})$ by increasing the
$i$-th cell's level by one such that
\[e_{i}(\vec{c})=v'-F_{d}(\vec{c}).\]
For good rewriting
performance, it is beneficial to make the update diversity of cell
states large.

\begin{lemma}\label{lemma:1}
Let $\vec{c}=(c_{1},c_{2},\dots,c_{n})$ be a cell state where for
$i=1,2,\dots,n$, $c_{i}\in \{0,1,\dots,q-2\}$. With the rewriting
code of Construction~\ref{constr:01}, the update diversity of
$\vec{c}$ is
\[\abs{\mathset{\theta_{w(\vec{c}),i} ~|~ i=1,2,\dots,n}}.\]
\end{lemma}
\begin{IEEEproof}
For $i=1,2,\dots,n$, we have
\begin{align*}
e_{i}(\vec{c}) & = F_{d}(N_{i}(\vec{c})) - F_{d}(\vec{c}) \\
& = \sum_{j=1}^{n}\theta_{w(\vec{c}),j}c_{j}+
\theta_{w(\vec{c}),i} +\sum_{j=0}^{w(\vec{c})}a_{j} \\
& \phantom{=} -
\sum_{j=1}^{n}\theta_{w(\vec{c})-1,j}c_{j}-\sum_{j=0}^{w(\vec{c})-1}a_{j} \\
& =  \theta_{w(\vec{c}),i} + a_{w(\vec{c})} +
\sum_{j=1}^{n}(\theta_{w(\vec{c}),j}-\theta_{w(\vec{c})-1,j})c_{j}
\end{align*}
Only the first term, $\theta_{w(\vec{c}),i}$, depends on $i$.
Hence the update diversity of $\vec{c}$ is
\[\abs{\mathset{e_{i}(\vec{c}) ~|~ i=1,2,\dots,n}}=
\abs{\mathset{\theta_{w(\vec{c}),i} ~|~ i=1,2,\dots,n}}.\]
\end{IEEEproof}

Therefore, to make the update diversity of cell states large, we
can make $\theta_{w(\vec{c}),1}, \theta_{w(\vec{c}),2}, \dots,
\theta_{w(\vec{c}),n}$ take as many different values as possible.
A simple solution is to let $\theta_{w(\vec{c}),i}=i$ for
$i=1,2,\dots,n$.

\subsection{Robustness}
In the following, we present our code for $n \geq L$. (The case of smaller $n$ can be dealt with using Construction~\ref{construction:5}.) The code uses randomness in the code construction to combat adversarial rewrite sequences. We then analyze the asymptotic optimality of the code for $nq \geq  L \log{L}$, and show that it optimizes the constant in the asymptotic performance to $1-\varepsilon$.

For $i=1,2,\dots,L$, we
define
\[g_{i}=\mathset{j ~|~ 1 \le j \le n,j\equiv i \pmod{L}}.\] For
example, if $n=8,L=3$, then
$g_{1}=\mathset{1,4,7},g_{2}=\mathset{2,5,8},g_{3}=\mathset{3,6}$.
For $i=1,2,\dots,L$, $\abs{g_{i}}$ is either $\floorenv{n/L}$
or $\ceilenv{n/L}$. We define \[h_{i}=\sum_{j\in
g_{i}}c_{j},\] where $c_{j}$ is the $j$-th cell's level. For
$i=1,2,\dots,L$, we have \[h_{i}\in \{0,1,\dots,|g_{i}|(q-1)\}.\]
We consider $g_{i}$ as a \emph{super cell} whose \emph{level} is
$h_{i}$.

\begin{construction}\label{constr:02}
(\textsc Robust Code)

For $i=0,1,\dots, n(q-1)-1$, choose the parameter $a_{i}$
independently and uniformly randomly from the set
$\mathset{0,1,\dots,L-1}$.

We define a randomized rewriting code $\cC_\cQ$ by its decoding function
\begin{equation}
\label{eq:strongdecode}
F_{d}(\vec{c})=\sum_{i=1}^{L}ih_{i}+\sum_{i=0}^{w(\vec{c})-1}a_{i}.
\end{equation}
By default, if $\vec{c}=(0,0,\dots,0)$, then $F_{d}(\vec{c})=0$.
When rewriting the data, we take the same greedy approach as in
Construction~\ref{constr:01}. \hfill $\Box$
\end{construction}

When we consider $g_{1},g_{2},\dots,g_{L}$ as $L$ ``super cells''
whose levels are $\vec{c'}=(h_{1},h_{2},\dots,h_{L})$, we have
\[w(\vec{c})=\sum_{i=1}^{n}c_{i}=\sum_{i=1}^{L}h_{i}=w(\vec{c'}).\]
The code of Construction~\ref{constr:02} may be seen as a rewriting code
that stores the data of alphabet size $L$ in $L$ super cells,
whose decoding function is \eqref{eq:strongdecode}.
Each of the super cells has either $(q-1)\floorenv{n/L}+1$
levels or $(q-1)\ceilenv{n/L}+1$ levels.

\begin{lemma}\label{lemma:3}
Let $\vec{c'}=(h_{1},h_{2},\dots,h_{L})$ be a super-cell state
where for $i=1,2,\dots,L$, $h_{i} \le (q-1)\floorenv{n/L}
-1$. With the rewriting code of Construction~\ref{constr:02}, the
update vector of the super-cell state $\vec{c'}$ is
\[u(\vec{c'})=\parenv{1+a_{w(\vec{c'})},2+a_{w(\vec{c'})},\dots,L+a_{w(\vec{c'})}},\]
and the update diversity of the super-cell state $\vec{c'}$ is $L$.
\end{lemma}
\begin{IEEEproof}
For $i=1,2,\dots,L$,
$N_{i}(\vec{c'})=(h_{1},\dots,h_{i-1},h_{i}+1,h_{i+1},\dots,h_{L})$,
so
\[e_{i}(\vec{c'}) = F_{d}(N_{i}(\vec{c'})) - F_{d}(\vec{c'})
= i+ a_{w(\vec{c'})}\]
and we get the conclusions.
\end{IEEEproof}

Therefore, if the current super-cell state is
$\vec{c'}=(h_{1},h_{2},\dots,h_{L})$ where for $i=1,2,\dots,L$,
$h_{i} \le (q-1)\floorenv{n/L}-1$, for the next rewrite, we
only need to increase one super-cell's level by one (which is
equivalent to increasing one flash-memory cell's level by one).

\begin{lemma}\label{lemma:2}
Let $\vec{c'}=(h_{1},h_{2},\dots,h_{L})$ be a super-cell state
where for $i=1,2,\dots,L$, $h_{i} \le (q-1)\floorenv{n/L}
-1$. With the rewriting code of Construction~\ref{constr:02}, if
$\vec{c'}$ is the current super-cell state, then no matter which
value the next rewrite changes the data to, the next rewrite will
only increase one super cell's level by one, and this super cell
is uniformly randomly selected from the $L$ super cells. What is
more, the selection of this super cell is independent of the past
rewriting history (that is, independent of the super cells whose
levels were chosen to increase for the previous rewrites).
\end{lemma}
\begin{IEEEproof}
Let $\vec{c'}$ be the current super-cell state, and assume the
next rewrite changes the data to $v'$. By Lemma~\ref{lemma:3}, we
will realize the rewrite by increasing the $i$-th super cell's
level by one such that $i+a_{w(\vec{c'})}=v'-F_{d}(\vec{c'})$.
Since the parameter $a_{w(\vec{c'})}$ is uniformly randomly chosen
from the set $\{0,1,\dots,L-1\}$, $i$ has a uniform random
distribution over $\{1,2,\dots,L\}$.

The same analysis holds for the previous rewrites. Note that with
every rewrite, the weight of the super cells, $w(\vec{c'})$,
increases. Since $a_{0},a_{1},\dots,a_{n(q-1)-1}$ are i.i.d.
random variables, the selection of the super cell for this rewrite
is independent of the selection for the previous rewrites.
\end{IEEEproof}

The above lemma holds for every rewrite sequence. We now prove
that the randomized rewriting code of Construction~\ref{constr:02}
is \emph{robust}.

\begin{theorem}\label{theorem:3}
Let $\cC_\cQ$ be the randomized rewriting code of Construction~\ref{constr:02}.
Let $\vec{v}=(v_{1},v_{2},v_{3},\dots)$ be any rewrite sequence.
For any constant $\varepsilon>0$ there exists a constant $c=c(\varepsilon)>0$ such that if $nq \geq cL\log{L}$,
then \[\E(t(\cC_\cQ|\vec{v}))\geq(1-\varepsilon)n(q-1),\]
and therefore $\cC_\cQ$ is a robust code.
\end{theorem}

\begin{IEEEproof}
Consider $L$ bins such that the $i$-th bin can hold
$(q-1)\abs{g_{i}}$ balls.
We use $h_{i}$ to denote the number of balls in the $i$-th bin.
Note that every bin can contain at least
$(q-1)\cdot \floorenv{\frac{n}{L}}$ balls and at most
$(q-1)\cdot \ceilenv{\frac{n}{L}}$ balls.  By
Lemma~\ref{lemma:2}, before any bin is full, every rewrite throws a
ball uniformly at random into one of the $L$
bins, independently of other rewrites.
The rewriting process can always continue before any bin
becomes full. Thus, the number of rewrites supported by the code
$\cC_\cQ$ is at least the number of balls thrown to make at least one
bin full.

Suppose that
$n(q-1)-\alpha\sqrt{nq}$ balls are independently and uniformly at random thrown into
$L$ bins, and there is no limit on the capacity of any bin. Here, we set
$\alpha$ to be $c\sqrt{L\log{L}}$ for a sufficiently large constant $c$. For $i=1,2,\dots,L$, let $x_{i}$ denote the
number of balls thrown into the $i$-th bin. Clearly,
\[\E(x_{i})=\frac{n(q-1)}{L}-\frac{\alpha\sqrt{nq}}{L}.\] By the
Chernoff bound,
\[\pr\parenv{x_i\geq (q-1)\cdot \floorenv{\frac{n}{L}}}
\leq e^{-\Omega(\alpha^2/L)}.\]
By the union bound, the probability that one or more of the $L$
bins contain at least $(q-1)\cdot \floorenv{\frac{n}{L}}$
balls is therefore upper bounded by $Le^{-\Omega(\alpha^2/L)}$. By our setting of $\alpha$ we have
$Le^{-\Omega(\alpha^2/L)}=2^{-\Omega(c^2)}$.

Therefore, when $n(q-1)-\alpha\sqrt{nq}$ balls are independently and
uniformly at random thrown
into $L$ bins, with high probability, all the $L$ bins have
$(q-1)\cdot \floorenv{\frac{n}{L}}-1$ or fewer balls. This suffices to conclude our assertion.
Notice that our proof implies that {\em with high probability} (over $\cQ$) the value of $t(\cC_\cQ|\vec{v})$ will be large.
This stronger statement implies the asserted one in which we consider $E(t(\cC_\cQ|\vec{v}))$.
\end{IEEEproof}

\section{Concluding Remarks}\label{section:conclusions}

In this paper, we presented a flexible rewriting model that
generalizes known rewriting models, including those used by WOM codes,
floating codes and buffer codes. We presented a novel code
construction, the trajectory code, for this generalized rewriting
model and proved that the code is asymptotically optimal for a very
wide range of parameter settings, where the performance is measured by
the number of rewrites supported by flash-memory cells in the worst
case. We also studied the expected performance of rewriting codes, and
presented a randomized robust code. It will be interesting to apply
these new coding techniques to wider constrained-memory applications,
and combine rewriting codes with error correction.  These remain as
our future research topics.

\bibliographystyle{IEEEtranS}

\end{document}